\documentclass[conference]{IEEEtran}
\IEEEoverridecommandlockouts
\usepackage{cite}
\usepackage{amsmath,amssymb,amsfonts}
\usepackage{algorithmic}
\usepackage{graphicx}
\usepackage{caption}
\usepackage{textcomp}
\usepackage{booktabs}
\usepackage{subcaption}
\usepackage{fancyhdr}
\usepackage{xcolor}
\def\BibTeX{{\rm B\kern-.05em{\sc i\kern-.025em b}\kern-.08em
    T\kern-.1667em\lower.7ex\hbox{E}\kern-.125emX}}
    
\usepackage{amssymb}

\fancypagestyle{myfooter}{
  \fancyhf{} 
  
  \fancyfoot[C]{\small{\textit{Published as} A. Trisovic, "Cluster Analysis of Open Research Data and a Case for Replication Metadata," 2022 IEEE 18th International Conference on e-Science (e-Science), Salt Lake City, UT, USA, 2022, pp. 423-424, doi: 10.1109/eScience55777.2022.00069.}} 
}
\fancypagestyle{plain}{%
  \fancyhf{} 
  
  \fancyfoot[C]{\small{\textit{Published as} A. Trisovic, "Cluster Analysis of Open Research Data and a Case for Replication Metadata," 2022 IEEE 18th International Conference on e-Science (e-Science), Salt Lake City, UT, USA, 2022, pp. 423-424, doi: 10.1109/eScience55777.2022.00069.}} 
}

\fancypagestyle{plain}{
  \fancyhf{} 
  \fancyfoot[C]{\thepage} 
}

\begin{document}

\title{Cluster Analysis of Open Research Data and a Case for Replication Metadata}

\author{\IEEEauthorblockN{Ana Trisovic}
\IEEEauthorblockA{\textit{Harvard Biostatistics \&} \textit{IQSS} \\
\textit{Harvard University}\\
Cambridge, MA, USA \\
anatrisovic@g.harvard.edu}
}

\maketitle
\pagestyle{myfooter}

\begin{abstract}
Research data are often released upon journal publication to enable result verification and reproducibility. For that reason, research dissemination infrastructures typically support diverse datasets coming from numerous disciplines, from tabular data and program code to audio-visual files. Metadata, or data about data, is critical to making research outputs adequately documented and FAIR. Aiming to contribute to the discussions on the development of metadata for research outputs, I conducted an exploratory analysis to determine how research datasets cluster based on what researchers organically deposit together. I use the content of over 40,000 datasets from the Harvard Dataverse research data repository as my sample for the cluster analysis. I find that the majority of the clusters are formed by single-type datasets, while in the rest of the sample, no meaningful clusters can be identified. For the result interpretation, I use the metadata standard employed by DataCite, a leading organization for documenting a scholarly record, and map existing resource types to my results. About 65\% of the sample can be described with a single-type metadata (such as Dataset, Software or Report), while the rest would require aggregate metadata types. Though DataCite supports an aggregate type such as a Collection, I argue that a significant number of datasets, in particular those containing both data and code files (about 20\% of the sample) would be more accurately described as a Replication resource metadata type. Such resource type would be particularly useful in facilitating research reproducibility.
\end{abstract}


\section{Introduction}

Computational research across the sciences can be vastly diverse. Researchers use multiple types of file formats to record data in their studies, from numerical to audio-video recordings. They also use different software applications or programming languages to analyze this data. Upon completing a study, the data, documents and code resources are often released online for verification, reuse and reproducibility purposes. To maximize the opportunities for efficient discovery and reuse of the research resources, they are deposited on the dissemination infrastructures that provide visibility on the web and comply with the FAIR principles. FAIR principles~\cite{wilkinson2016fair} have emerged as guidelines to facilitate making digital research resources findable, accessible, interoperable, and reusable. They have been widely recognized and adopted as the vision for research infrastructures supporting effortless data reuse. In practice, this is achieved with metadata – data about data or adequate and machine-actionable documentation of the shared resources~\cite{jacobsen2020fair}. 

In practice, however, the research practices are constantly evolving, with the use of ever-increasing data volume, compute and complex methods, which leads to new challenges in metadata implementation and its support in research repositories. This paper presents an exploratory data analysis of the open research data and metadata published on Harvard Dataverse, an open-source data repository platform for sharing, archiving, and citing research data. The sample for the study is the content of over 40,000 open research datasets, including over 500,000 files. The study is intended to facilitate open data sharing and reuse by providing insight into research data clusters, which could be valuable for metadata developments. It should be of interest to researchers, digital libraries and research infrastructure communities across the sciences.

\begin{figure}[h]
    \centering
    \includegraphics{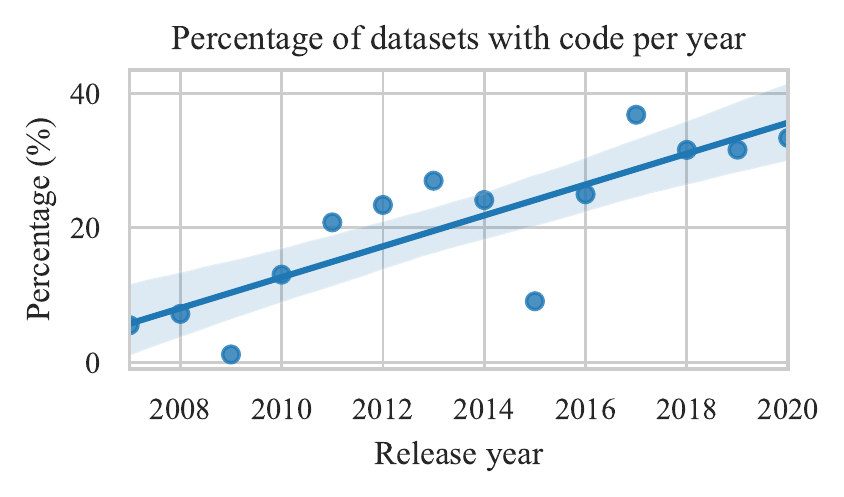}
    \caption{Percentage of datasets published at Harvard Dataverse repository that contain code files per year. Sample size: 40,634 datasets.}
    \label{fig:code}
\end{figure}

\section{Methods and results}

Harvard Dataverse repository implements FAIR principles and captures metadata on all deposited datasets. Each dataset contains files that are identified with a media type (also known as a \textit{mime} type), which is a two-part label used to identify file formats on the web. On Dataverse repositories, these labels are used for object handling, like its preview in the browser. There are over 300 different media types on Harvard Dataverse, and about 200 of them were mapped into objects (facets) as a part of data preparation and cleaning steps. After media types were mapped into objects in each dataset, the sample size counted 40,634 dataset entries with 586,169 files (fully mapped as objects) from the initial count of about 45,000 datasets. All dataset entries that had unknown object types were removed from the analysis. The sample is formatted such that each Dataverse dataset contains a count of its objects (archive, audio, code, data, document, image, shape, text, video), and as such is used for further analysis. The initial analysis suggests that most datasets contain only a few types of objects and that the portion of datasets that contain code has been increasing over the years (Fig.~\ref{fig:code}). The small number of objects per dataset may suggest that only a few object types are frequently shared together, such as data and code or data and documentation. We can expect to see single-object datasets or code objects grouped with text objects (as their documentation). Examining cluster tendency and clustering in the sample can help us identify optimal dataset metadata based on what objects researchers organically deposit together.

\begin{figure}[htbp]
    \centering
    \includegraphics{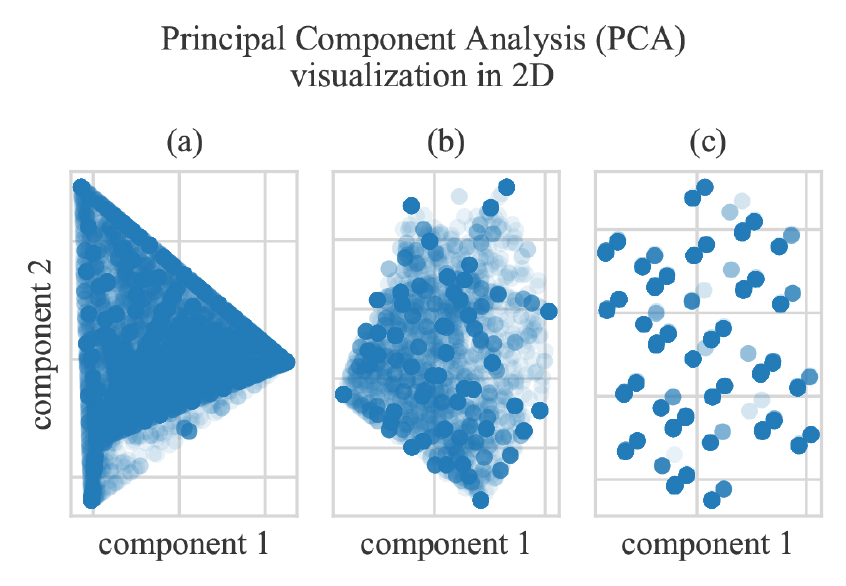}
    \caption{(a) original dataset, (b) original reduced dataset, and (c) binary dataset.}
    \label{fig:pca}
\end{figure}

Clustering is a process where natural groupings within a set are determined, such that the items in each group exhibit more similarity than items in other groups. The Hopkins statistic~\cite{lawson1990new} and the multimodality tests were used to evaluate the clustering tendency of the sample. The Hopkins statistic measures the spatial randomness of the data by comparing it to generated sample with uniform distribution and returns the probability that the data has a uniform random distribution. The Hopkins statistic score is between 0 and 1, where values close to 0 can be interpreted as high cluster tendency, and values above 0.3 express no clustrerability. We obtain a score of $h = 0.0026$,  suggesting that the dataset is highly clusterable, though the test does not ensure there is more than one cluster. The dip test~\cite{hartigan1985dip} is a widely used multimodality test that computes a statistic called the dip, which is defined as the maximum distance between the empirical distribution and the closest uniform distribution. It returns a \textbf{p}-value as the probability of observing the input being generated from a unimodal distribution (its null hypothesis). If only a single mode is present, the \textbf{p}-value will be large, suggesting that the data cannot be clustered. A small p-value, such as ours $p \leq 0.001$ (the dip value of $0.12$) suggests that multiple modes (and multiple clusters) are present. 

Principal component analysis (PCA) is an unsupervised learning approach used in exploratory analyses that reduces data from high dimensions to lower dimensions while preserving the covariance in the data. The dimensionality reduction can help in visualizing high-dimensional datasets and intuitively judging whether they have meaningful clusters. The scatter plots of the data in two dimensions are shown in Fig.~\ref{fig:pca}. The explained variance ratio is the percentage of covariance explained by the reduced dataset. In this case, they are 51\%, 61\%, and 66\%. This means that the sample variance is not fully preserved in two dimensions. The sample is used in PCA and plotted for a higher number of dimensions (up to five), but no clusters were observed.

\section{Discussion and outlook}

The study results suggest that the sample is highly clusterable, but at the same time, it cannot be "cleanly clustered". Though perplexing, such a result can be interpreted to contain discrete clusters, but those are mostly single-type clusters (65\% of the sample), while the rest of the sample forms a cluster that cannot be further meaningfully clustered. In the other words, most datasets can be described as a single-type metadata, but for the rest, an aggregate type that incorporates various research objects is needed. For the aggregate cluster, a metadata type such as \textit{Replication metadata} or other flexible metadata (that bundles different types of objects) would be useful.

Using widely recognized DataCite Metadata Schema (datacite.org), we can further interpret these results. Currently, the DataCite Metadata Schema recognizes 28 metadata types (\texttt{resourceTypeGeneral}), including \textit{Audiovisual},  \textit{Collection},  \textit{Dataset},  \textit{Software},  \textit{Text}. Our sample could be transformed to fit the DataCite metadata description in the following way: 37\% (14859) can be classified under the \textit{Dataset} type, 11\% (4296) as \textit{Report}, \textit{Preprint}, \textit{Journal Article} or \textit{Dissertation}, 0.1\% (73) would be classified as \textit{Audiovisual}, 1\% (378) as \textit{Software}, 10\% (4101) as \textit{Image} and 7\% (2977) as \textit{Text}. For the rest of the sample (20\% (8145) containing code files and 14\% (5805) without code files), an aggregate metadata type would be optimal. For instance, a \textit{Collection} resource type implies that the object contains various elements and could be used in this case, though it may be a vague description for many of the datasets. A significant portion of the aggregate cluster (20\%) contains datasets that contain research software or code, which are among the most fragile (software- and system-dependent) research artefacts and need to be shared according to specific guidelines. With \textit{Replication metadata}, a published research study would potentially be better documented, easier to grasp and reproduce as all its resources would be gathered in a single bundle, including data, code, documentation, slides, and review, thus decreasing the chance of scattered reporting.

\bibliographystyle{IEEEtran}
\bibliography{references}

\begin{thebibliography}{1}
\providecommand{\url}[1]{#1}
\csname url@samestyle\endcsname
\providecommand{\newblock}{\relax}
\providecommand{\bibinfo}[2]{#2}
\providecommand{\BIBentrySTDinterwordspacing}{\spaceskip=0pt\relax}
\providecommand{\BIBentryALTinterwordstretchfactor}{4}
\providecommand{\BIBentryALTinterwordspacing}{\spaceskip=\fontdimen2\font plus
\BIBentryALTinterwordstretchfactor\fontdimen3\font minus
  \fontdimen4\font\relax}
\providecommand{\BIBforeignlanguage}[2]{{%
\expandafter\ifx\csname l@#1\endcsname\relax
\typeout{** WARNING: IEEEtran.bst: No hyphenation pattern has been}%
\typeout{** loaded for the language `#1'. Using the pattern for}%
\typeout{** the default language instead.}%
\else
\language=\csname l@#1\endcsname
\fi
#2}}
\providecommand{\BIBdecl}{\relax}
\BIBdecl

\bibitem{wilkinson2016fair}
M.~D. Wilkinson, M.~Dumontier, I.~J. Aalbersberg, G.~Appleton, M.~Axton,
  A.~Baak, N.~Blomberg, J.-W. Boiten, L.~B. da~Silva~Santos, P.~E. Bourne
  \emph{et~al.}, ``The fair guiding principles for scientific data management
  and stewardship,'' \emph{Scientific data}, vol.~3, no.~1, pp. 1--9, 2016.

\bibitem{jacobsen2020fair}
A.~Jacobsen, R.~de~Miranda~Azevedo, N.~Juty, D.~Batista, S.~Coles, R.~Cornet,
  M.~Courtot, M.~Crosas, M.~Dumontier, C.~T. Evelo \emph{et~al.}, ``Fair
  principles: interpretations and implementation considerations,'' 2020.

\bibitem{lawson1990new}
R.~G. Lawson and P.~C. Jurs, ``New index for clustering tendency and its
  application to chemical problems,'' \emph{Journal of chemical information and
  computer sciences}, vol.~30, no.~1, pp. 36--41, 1990.

\bibitem{hartigan1985dip}
J.~A. Hartigan, P.~M. Hartigan \emph{et~al.}, ``The dip test of unimodality,''
  \emph{Annals of statistics}, vol.~13, no.~1, pp. 70--84, 1985.

\end{thebibliography}

\end{document}